\documentclass[journal,onecolumn]{IEEEtran}

\usepackage[utf8]{inputenc}
\usepackage{graphicx}
\usepackage{amsmath,amssymb}
\usepackage{enumitem}
\usepackage{booktabs}
\usepackage{multirow}

\begin{document}

\title{Virtualizing RAN: Science, Strategy, and Architecture of Software‑Defined Mobile Networks}

\author{\IEEEauthorblockN{Ryan Barker} \\
    \IEEEauthorblockA{Holcombe Department of Electrical and Computer Engineering, \\
    Clemson University, Clemson, SC, USA \\
    Email rcbarke@clemson.edu}
}

\maketitle

\begin{abstract}
Virtualizing the Radio–Access Network (RAN) is increasingly viewed as an
enabler of affordable\,5G expansion and a stepping-stone toward AI-native\,6G.
Most discussions, however, still approach spectrum policy, cloud
engineering and organizational practice as separate topics.  This paper
offers an integrated perspective spanning four pillars—\emph{science,
technology, business strategy} and \emph{culture}.  A comparative U.S.\ case
study illustrates how mid-band contiguity, complemented by selective
mmWave capacity layers, can improve both coverage and churn when orchestrated
through software-defined carrier aggregation.  We derive analytic capacity and
latency bounds for Split\,7.2$\times$ vRAN/O-RAN deployments, quantify the
throughput penalty of end-to-end 256-bit encryption, and show how GPU/FPGA
off-load plus digital-twin–driven automation keeps the hybrid-automatic-repeat
request (HARQ) round-trip within a 0.5\,ms budget.  When these technical
enablers are embedded in a physics-first delivery roadmap, average vRAN
cycle time drops an order of magnitude—even in the presence of cultural
head-winds such as “dual-ladder’’ erosion.  Three cybernetic templates—the
\emph{Clock-Hierarchy Law}, \emph{Ashby’s Requisite Variety} and a
\emph{delay–cost curve}—are then used to explain why silo-constrained
automation can amplify, rather than absorb, integration debt.  Looking
forward, silicon-paced 6G evolution (9–12-month node shrinks, sub-THz
joint communication-and-sensing\footnote{JCAS = Joint Communication and
Sensing.}, chiplet architectures and optical I/O) calls for a
\emph{dual-resolution planning grid} that couples five-year spectrum physics
with six-month “silicon sprints.’’  The paper closes with balanced,
action-oriented recommendations for operators, vendors and researchers on
sub-THz fronthaul, AI-native security, energy-proportional accelerators and
zero-touch assurance.
\end{abstract}

\begin{IEEEkeywords}
Virtualized RAN, Open RAN, Split 7.2 ×, Open RAN, Digital Twin, Carrier Aggregation, Clock-Hierarchy Law, Requisite Variety, SDN, AI-native Security, Zero-Trust, Change Management, 5G, 6G.
\end{IEEEkeywords}

\section{Introduction}
\label{sec:intro}

The ﬁfth‑generation (5G) mobile era has amplified long‑standing frictions between
radio‑frequency engineering, real‑time computing and business execution.
On the technical front, ever‑rising traffic, sub‑millisecond latency targets and
energy constraints expose the limits of proprietary, hardware‑centric
base‑station designs.
Concurrently, the commercial success of low‑/mid‑band spectrum plays such as
T‑Mobile’s “Layer Cake’’ contrasts with the mixed customer experience of
mmWave‑heavy roll‑outs, highlighting an urgent need to balance coverage,
capacity and cost~\cite{tmobile_layercake,verizon_cband}.
Open and virtualized Radio Access Networks (vRAN/O‑RAN) promise to break this
impasse by disaggregating radio functions, embracing software‑defined
networking (SDN) principles and enabling AI‑assisted automation
\cite{Understanding_ORAN,oran_spec}. Yet, their adoption is throttled by cultural inertia inside incumbent local
exchange carriers (ILECs), the shift of technical career ladders\footnote{The views expressed in this work are solely the author’s and do not reflect those of any current or former employer.}, and open
security questions spanning 5G cipher downgrade attacks to side‑channel
leakage~\cite{never_let_me_down_again,snow_sca}.

\subsection{Problem Context and Motivation}
\textit{Cost context—}Early commercial vRAN roll-outs such as
Rakuten Mobile report a hardware bill-of-materials of roughly
\$30\,k per open-RAN radio unit (RU) versus \$100\,k for a like-for-like
legacy macro swap, translating into a $\approx 70\%$ CapEx reduction and
$\approx 40\%$ lower three-year Opex when pooled compute is factored
in~\cite{rakuten_bom}.

\begin{itemize}
\item \textbf{Hardware lock‑in vs.\ software agility.}
  Legacy base‑band units (BBUs) and custom accelerators excelled in
  LTE, but struggle to scale for multi‑band carrier aggregation, Massive‑MIMO
  and network slicing without prohibitive cost or power
  draw~\cite{keysight_mimo,ericsson_ca}.
\item \textbf{Spectrum scarcity and uneven asset portfolios.}
  National coverage requires low‑frequency reliability while urban hot‑spots
  crave mmWave capacity; auction outcomes create fragmented holdings that
  only software‑defined orchestration can exploit efficiently~\cite{fcc_auction110}.
\item \textbf{Operational complexity.}
  Disaggregated O‑RAN stacks add dozens of new interfaces (E2,O1,A1,~\textit{etc.}),
  shifting the bottleneck from hardware integration to real‑time coordination
  and security of multi‑vendor components~\cite{securing_oran}.
\item \textbf{Cultural and governance barriers.}
  Kotter‑style change errors, outsourcing bias and weakened technical career
  paths stall vRAN adoption despite clear Total‑Cost‑of‑Ownership (TCO)
  advantages~\cite{openranpolicy,cisco_son}.
\end{itemize}

These converging pressures motivate a holistic investigation that spans
radio physics, SDN architecture, AI/ML control loops, security engineering and
organizational strategy.

\subsection{Scope and Contributions}

\vspace{-0.25em}
This paper offers a cross‑disciplinary commentary and technical review aimed at
researchers, practitioners and policy makers who shape next‑generation mobile
networks:

\begin{enumerate}
  \item \textbf{Taxonomy \& Reference Model}—I codify the functional splits,
        timing budgets and security primitives that underpin Cloud‑RAN,
        vRAN and O‑RAN deployments, bridging 3GPP, O‑RAN Alliance and SDN
        nomenclature.
  \item \textbf{Spectrum–Compute Co‑Design}—I derive an analytic framework
        that links carrier‑aggregation scenarios, BBU/DU processing ceilings
        and GPU/FPGA acceleration options to real‑world power budgets.
  \item \textbf{AI‑Native Automation Blueprint}—Building on open testbeds
        such as OAIC and X5G, I map how Digital‑Twin pipelines and RIC
        xApps close the loop between planning, optimization and
        self‑healing~\cite{oaic,x5g,dApps}.
  \item \textbf{Security Cost Model}—Extending recent measurements
        \cite{securing_oran}, I quantify the latency/throughput trade‑offs
        of 256‑bit cipher enforcement (SNOW‑V, AES‑GCM, ZUC‑256) across open
        interfaces.
  \item \textbf{Change‑Management Analysis}—I relate technical road‑blocks to
        organizational mis‑alignments, illustrating lessons from Verizon’s
        mmWave pivot and T‑Mobile’s mid‑band momentum.
\end{enumerate}

\subsection{Paper Organization}

\vspace{-0.25em}
Section~\ref{sec:landscape} surveys spectrum economics, auction dynamics and
operator strategies.
Section~\ref{sec:tech_found} drills into the radio and compute foundations of
virtualized RAN, including carrier‑aggregation mathematics and O‑RAN logical
interfaces.
Section~\ref{sec:enablers} explores SDN control, edge acceleration and AI/ML
pipelines, while highlighting security implications.
Section~\ref{sec:cases} distils empirical lessons from U.S. 5G deployments, change‑management literature, and the author's career.

\vspace{0.6em}
\noindent
\textbf{Positioning within Systems Theory—}
Having drilled down to FFT-symbol deadlines and carrier–aggregation knapsacks, the
remainder of the paper deliberately zooms \emph{out} to the macro–dynamics that
govern how large organizations absorb (or repel) exponential technology.  
Section~\ref{sec:systems} frames virtualized RAN as an unfolding
\emph{socio-technical control problem}: the \emph{Clock-Hierarchy Law} shows how
strategy outruns execution when internal R\&D clocks are allowed to stall;
\emph{Ashby’s Requisite Variety} quantifies the resilience lost when dual-career
ladders rust; and the \emph{delay–cost curve} explains why local automation
without a unifying road-map creates compounding integration debt.  
\emph{Culture Eats Strategy for Breakfast: The Nuance of Technology and
Infrastructure}—adds balance by contrasting “up-or-out’’ M\&A playbooks with
pockets of deep technical practice that still thrive inside the same firms,
showing that the issue is less \emph{which} company than \emph{how} clocks,
careers, and capital are synchronized.  
Taken together, these cybernetic lenses turn the paper from a “how-to’’ guide
for vRAN deployment into a theory-backed playbook for scaling AI-native 6G
without replaying 5G’s cultural mis-steps.  
Readers focused on RF and compute may stop at
Section~\ref{sec:enablers}; those charged with steering national networks
through the next decade should continue into Section~\ref{sec:systems}.  
Finally, Section~\ref{sec:conclusion} synthesises the technical and cultural
threads and outlines open research directions toward an AI-native, zero-trust,
energy-neutral 6G.

\section{Industry Landscape: Spectrum Economics \& Strategic Postures}
\label{sec:landscape}

\subsection{Historical Foundations}

The organizational DNA of today’s mobile operators can be traced to the
pre‑divestiture \textit{Bell System} monopoly and its subsequent fragmentation
into Regional Bell Operating Companies (RBOCs) in 1984.
Those incumbents—henceforth \textit{Incumbent Local Exchange Carriers
(ILECs)}—inherited a culture centred on engineering conservation, proprietary hardware,
five‑nine availability targets and top‑down capital allocation cycles.
While this model delivered unrivalled voice reliability, it also entrenched
long replacement cycles and vendor lock‑in that now clash with the
software‑driven ethos of virtualized RAN.
The deceleration of formal “dual‑career ladders’’ within many ILECs\footnote{A
dual ladder permits parallel technical and managerial advancement; its erosion
has diverted subject‑matter experts into project‑management roles, eroding
in‑house design competence.} further weakened
internal technology advocacy, tilting power toward cost‑optimized sourcing
teams and short‑term revenue goals.
Consequently, when 5G emerged, the U.S. incumbents reacted with hardware‑first
tactics—Verizon’s mmWave Ultra‑Wideband (UWB) Nationwide Coverage attempt being the canonical
example—rather than embracing software abstraction and agile spectrum use.

\subsection{Spectrum Assets and Auction Dynamics}

\begin{figure}[!h]
    \centering
    \includegraphics[width=\columnwidth, ]{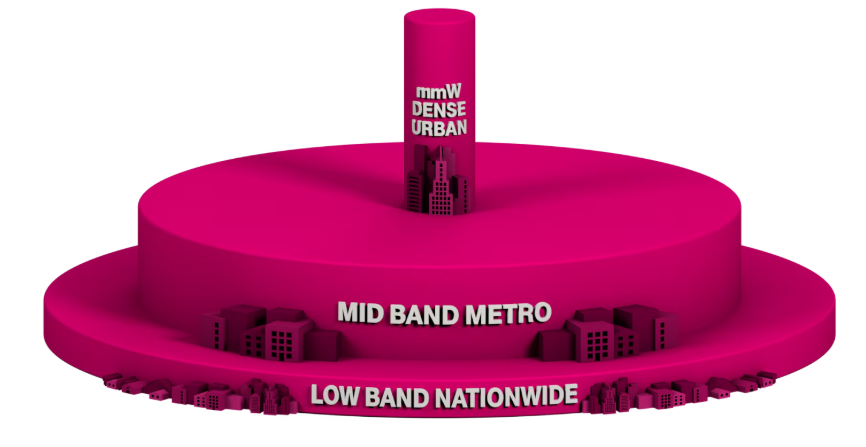} 
    \caption{T Mobile's balanced "Layer Cake'' 5G Deployment Strategy, emphasizing geographically-driven, right-sized spectrum deployment.}
    \label{fig:layercake}
\end{figure}

\paragraph*{Low, mid and high bands.}
Figure~\ref{fig:layercake} (T‑Mobile’s marketing visual) encapsulates the core
physics trade‑off:
\emph{low‑band} ($\leq$ 1 GHz) offers deep propagation and building penetration
but limited channel bandwidth;
\emph{mid‑band} (1–7.125 GHz) balances coverage and capacity, particularly in
the newly cleared 2.5 GHz, 3.45 GHz and C‑band (3.7–3.98 GHz) ranges;
\emph{high‑band/mmWave} (24–47 GHz) supplies extreme throughput and beamforming
gain at the expense of range and foliage loss.

\paragraph*{Auction chronology.}
U.S.\ spectrum is assigned via ascending‑clock auctions with partial reserve
prices; Table~\ref{tab:spectrum} summarizes headline events.
Notable milestones include Auction 97 (AWS‑3, \$44.9 B, 2015),
Auction 107 (C‑band, \$81 B, 2021) and Auction 110 (3.45 GHz, \$22.5 B,
2022)~\cite{fcc_auction110}.
Each round reshapes operator holdings and debt capacity, dictating where
carrier‑aggregation (CA) and Dynamic Spectrum Sharing (DSS) can be
deployed profitably.
While DSS briefly enabled 5G overlays in 700/850 MHz LTE carriers,
operators quietly de‑commissioned most deployments in 2024 after discovering a
20–40\% capacity penalty and scheduler complexity~\cite{lightreading_dss}.

\paragraph*{Spectrum-clearing timelines.}
C-band licences required a two-phase satellite relocation that stretched
from Dec 2020 (Phase 1, 46 PEAs) to Dec 2023 (Phase 2), delaying full-power
macro activation by 12–18 months after auction close.
Similarly, 3.45 GHz grants faced Department-of-Defense coordination
windows of $\approx 6$ months~\cite{fcc_clearing}.

\paragraph*{Contiguous vs.\ dis‑contiguous economics.}
Mid‑band licences are often fragmented across Partial Economic Areas (PEAs),
forcing operators either to bid aggressively for contiguous blocks or accept
smaller non‑contiguous channels that require additional hardware
resources~\cite{ericsson_ca}.
Virtualized DU/CU pools partially mitigate this by multiplexing disparate
carriers in software, but back‑to‑back spectrum blocks still command premium
prices due to lower per‑bit cost.

\begin{table}[!h]
  \centering
  \caption{Headline U.S. spectrum‑auction milestones (ascending‑clock format).}
  \label{tab:spectrum}
  \begin{tabular}{@{}lccc@{}}
    \toprule
    \textbf{Auction \#} & \textbf{Band / Service} & \textbf{Year} & \textbf{Gross Proceeds (USD B)} \\ \midrule
    97  & AWS‑3 (1.7/2.1 GHz) & 2015 & 44.9 \\
    107 & C‑band (3.7–3.98 GHz) & 2021 & 81.0 \\
    110 & 3.45 GHz Service & 2022 & 22.5 \\ \bottomrule
  \end{tabular}
\end{table}

\subsection{Operator Go‑to‑Market Strategies}

\paragraph*{T‑Mobile US —~“Layer Cake’’ playbook.}
The 2020 Sprint merger endowed T‑Mobile with 194 MHz of 2.5 GHz nationwide
spectrum, enabling an early‑mover mid‑band roll‑out.
Its three‑tier “Layer Cake’’—600 MHz for coverage, 2.5 GHz for capacity and
select mmWave in stadiums—delivered consistent user experience while keeping
CapEx in check~\cite{tmobile_layercake}.
Three‑carrier aggregation (600+2.5+n258) now exceeds 3 Gb/s peak
throughput in dense metros.

\paragraph*{Verizon — mmWave‐first pivot to C‑band.}
Leveraging vast 28/39 GHz holdings, Verizon launched UWB small‑cells to claim
multi‑Gb/s marketing headlines.
However, limited propagation demanded thousands of additional street‑level
poles; CapEx pressure and negative churn forced a pivot toward C‑band, where
Verizon secured 160–200 MHz in the top‑46 PEAs~\cite{verizon_cband}.
The firm is now re‑optimizing its network with Massive‑MIMO C‑band macro cells
supplemented by targeted mmWave FWA.

\paragraph*{AT\&T — Balanced but funding‑constrained.}
AT\&T split its bids between 3.45 GHz and C‑band while leveraging FirstNet
(700 MHz) commitments to densify low‑band coverage.
Financial leverage from WarnerMedia spin‑offs limits acceleration, making
RAN cost optimization a strategic imperative.

\subsection{Regulatory and Compliance Head‑winds}

\paragraph*{FAA radar altimeter dispute.}
The mid‑band roll‑out faced an unanticipated obstacle when the U.S.\ Federal
Aviation Administration raised concerns over 3.7–3.98 GHz emissions near
airport glide‑paths.
Temporary power‑down zones and antenna‑tilt constraints delayed commercial
activation in 50+ markets, illustrating how non‑technical agencies can sway
network timelines.

\paragraph*{NEPA, EME and historic‑preservation reviews.}
New macro or small‑cell sites require National Environmental Policy Act (NEPA)
screening and, where applicable, Tribal and historic‑structure consultation.
Processing times stretch from weeks (collocation on existing poles) to $>$ 12 months for green‑field towers, directly impacting annual build targets.

\paragraph*{Enforcement trends.}
The FCC has intensified enforcement of 911 location‑accuracy and EME rules:
Verizon’s \$0.95 M settlement in 2024 highlights financial exposure for
non‑compliance.
Open RAN deployments add further complexity because RU suppliers must certify
Part‑15 emissions independently, while the integrator remains liable for the
aggregate installation.

\paragraph*{Security mandates.}
Executive Order 13873 (“Securing the Information and Communications Technology
and Services Supply Chain’’) plus the Secure Networks Act restrict the use of
equipment from “covered’’ entities.
The Open RAN Policy Coalition argues that interface openness coupled with
U.S.‑based RIC/xApp ecosystems offers a path to both innovation and supply‑chain
resilience~\cite{openranpolicy}.
However, holistic zero‑trust postures—including 256‑bit cipher enforcement on
all open interfaces—remain a moving target given measurable performance costs
\cite{securing_oran}.

\vspace{0.3em}
\noindent
\textbf{Take‑away—}Spectrum economics, auction outcomes and
non‑technical regulation collectively shape the technical architecture and
timing of vRAN/O‑RAN adoption.  Operators that translate diverse holdings into
software‑defined, AI‑optimized capacity—while navigating compliance at
minimum Opex—are positioned to out‑scale hardware‑centric rivals.

\subsection{Replication evidence (IMC ’25).}
As this review was being finalized, Ghoshal \emph{et~al.} reported a coast‑to‑coast replication of the 2022 “on‑the‑wheels’’ study~\cite{ghoshal2022-wheels-replication}, measuring all three U.S.\ operators (Los Angeles$\to$Boston, Nov‑2024). They find that total 5G coverage improved to 49/94/45\% for Verizon/T‑Mobile/AT\&T, with T‑Mobile the only operator exhibiting extensive 5G~Standalone (60.5\%), while the others remain predominantly NSA. Median downlink throughput rose sharply (148/248/77~Mb/s), largely from higher mid‑band CA (2–4~CCs), but end‑to‑end latency barely moved (and even worsened for some technologies), reinforcing our claim that spectrum/compute upgrades alone do not cure RTT without transport and scheduling reform. The study also reports a practical retreat in high‑order MIMO usage (fewer 4$\times$4 deployments and layers exercised) and a higher HO rate (about 2.7–3.9 per mile) consistent with denser mid‑band footprints and NSA anchor effects. A head‑to‑head with Starlink shows near‑continuous availability and the lowest RTT, but much weaker UL and capped DL peaks, underscoring our emphasis on mid‑band contiguity plus SA as the scalable path to stable QoE~\cite{ghoshal2025-wheels-replication}.

\section{Technical Foundations of Radio Access Virtualization}
\label{sec:tech_found}

Virtualization reshapes the radio access network by decoupling time‑critical
baseband logic from monolithic, proprietary appliances.  
Before examining vRAN and O‑RAN abstractions I establish four technical
pillars: (i) the classical hardware chain, (ii) spectrum engineering
constraints that dictate cell‑site topology, (iii) the evolution from
centralized to virtualized RAN splits, and (iv) the service‑management
framework introduced by the O‑RAN Alliance.

\subsection{Canonical RAN Hardware Chain}
\label{ssec:canon_chain}

\begin{figure}[!h]
    \centering
    \includegraphics[width=0.5\columnwidth]{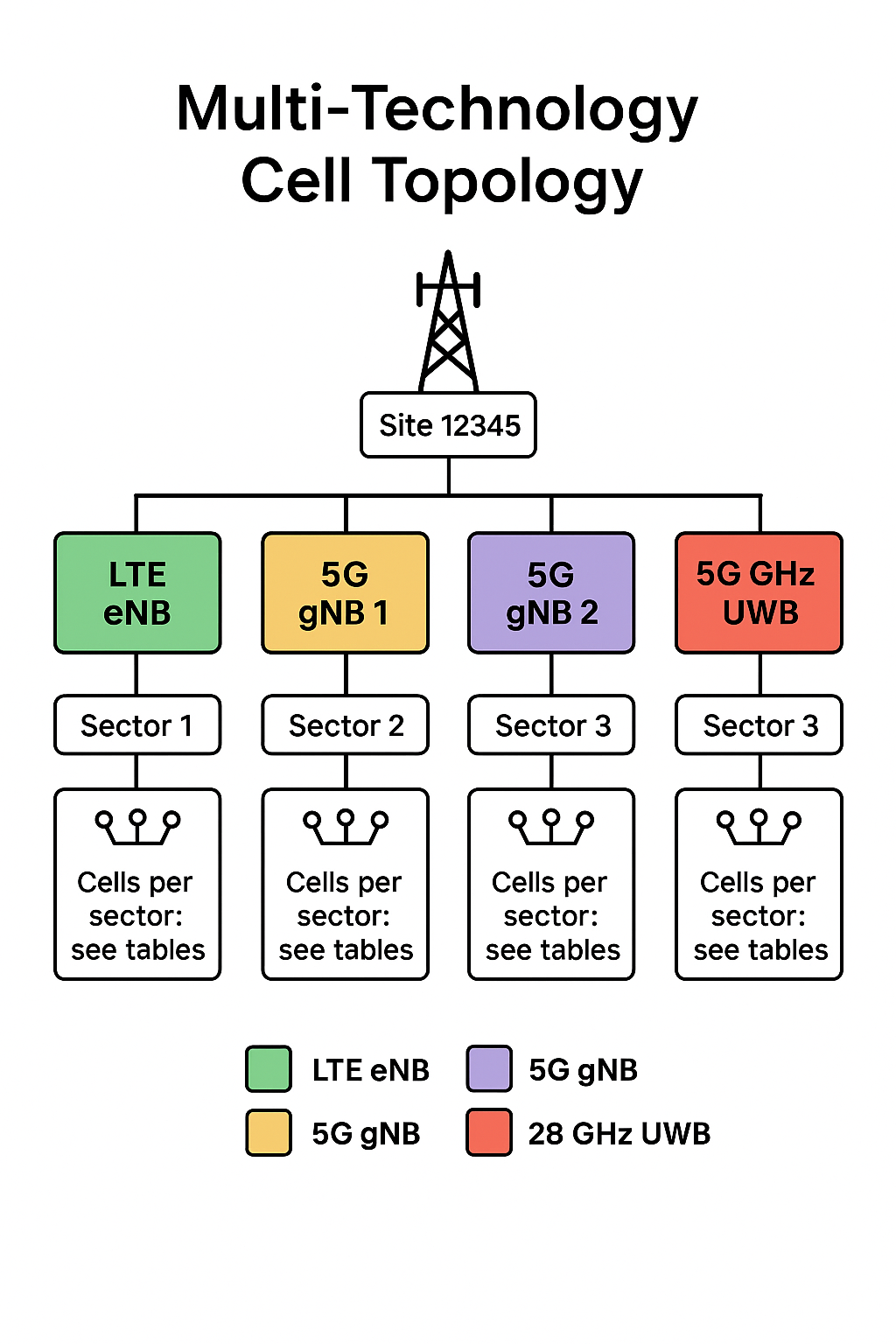} 
    \caption{Example cell numbering schema for a multi-technology deployment. Reference Tables \ref{tab:eNB152748}, \ref{tab:gNB1529748}, \ref{tab:gNB1529749}, \ref{tab:gNB1529750} for specific cell assignments.}
    \label{fig:cellTopology}
\end{figure}

A legacy macro eNB/gNB comprises four physical blocks:
\textbf{(1)~Cell‑Site Router (CSR)} for IP backhaul,
\textbf{(2)~Baseband or Digital Unit (BBU/DU)} that terminates
user‑plane (PDCP, RLC, MAC) and performs Layer‑1 processing,
\textbf{(3)~Radio Unit (RU)} which amplifies/filters I/Q streams, and
\textbf{(4)~Antenna array (ANT)} that radiates beams toward user equipment
(UE).  
Massive‑MIMO increases RU+ANT complexity: 64T64R arrays at C‑band add
$\approx{}20$ dB beamforming gain but incur 300–500 W incremental power
\cite{keysight_mimo}.  
Figure~\ref{fig:cellTopology} (adapted from a mock deployment architecture) shows how a single tower may host \textit{multiple} logical nodes:
two LTE eNBs plus three NR gNBs, each split into sectors (\texttt{\_1},
\texttt{\_2}, \texttt{\_3}) and per‑sector cells (\texttt{\_x\_y}).

\paragraph*{Capacity accounting.}
The DU firmware limits \emph{(i)}~total cell count (18–21 on
pre‑vRAN hardware) and \emph{(ii)}~cumulative \textit{Absolute Bandwidth}
(ABW) per DU (\(<\!160\) MHz FR1; \(>\!400\) MHz FR2 when split across
multiple component carriers).
When spectrum is highly fragmented, operators face a cellular “knapsack’’
problem: either deploy an extra DU or disable least‑profitable carriers to
stay within software limits.

\newpage

\begin{table}[htbp]
\centering
\caption{4G eNB 999748 – Sector/Cell–to–Frequency Mapping}
\label{tab:eNB152748}
\begin{tabular}{@{}lll@{}}
\toprule
\textbf{Sector} & \textbf{Cell ID} & \textbf{Frequency} \\
\midrule
\multirow{5}{*}{1}
  & 999748\_1\_1 & 700\,MHz \\
  & 999748\_1\_2 & AWS 2600\,MHz \\
  & 999748\_1\_3 & PCS 1900\,MHz \\
  & 999748\_1\_4 & PCS 2150\,MHz \\
  & 999748\_1\_5 & 850\,MHz \\
\midrule
\multirow{5}{*}{2}
  & 999748\_2\_1 & 700\,MHz \\
  & 999748\_2\_2 & AWS 2600\,MHz \\
  & 999748\_2\_3 & PCS 1900\,MHz \\
  & 999748\_2\_4 & PCS 2150\,MHz \\
  & 999748\_2\_5 & 850\,MHz \\
\midrule
\multirow{5}{*}{3}
  & 999748\_3\_1 & 700\,MHz \\
  & 999748\_3\_2 & AWS 2600\,MHz \\
  & 999748\_3\_3 & PCS 1900\,MHz \\
  & 999748\_3\_4 & PCS 2150\,MHz \\
  & 999748\_3\_5 & 850\,MHz \\
\bottomrule
\end{tabular}
\end{table}

\begin{table}[htbp]
\centering
\caption{5G gNB 9999748 – Sector/Cell–to–Frequency Mapping}
\label{tab:gNB1529748}
\begin{tabular}{@{}lll@{}}
\toprule
\textbf{Sector} & \textbf{Cell ID} & \textbf{Frequency} \\
\midrule
\multirow{4}{*}{1}
  & 9999748\_1\_2 & AWS 2600\,MHz \\
  & 9999748\_1\_3 & PCS 1900\,MHz \\
  & 9999748\_1\_4 & PCS 2150\,MHz \\
  & 9999748\_1\_5 & 850\,MHz \\
\midrule
\multirow{4}{*}{2}
  & 9999748\_2\_2 & AWS 2600\,MHz \\
  & 9999748\_2\_3 & PCS 1900\,MHz \\
  & 9999748\_2\_4 & PCS 2150\,MHz \\
  & 9999748\_2\_5 & 850\,MHz \\
\midrule
\multirow{4}{*}{3}
  & 9999748\_3\_2 & AWS 2600\,MHz \\
  & 9999748\_3\_3 & PCS 1900\,MHz \\
  & 9999748\_3\_4 & PCS 2150\,MHz \\
  & 9999748\_3\_5 & 850\,MHz \\
\bottomrule
\end{tabular}
\end{table}

\begin{table}[htbp]
\centering
\caption{5G gNB 9999749 – Sector/Cell–to–Frequency Mapping (C-Band)}
\label{tab:gNB1529749}
\begin{tabular}{@{}lll@{}}
\toprule
\textbf{Sector} & \textbf{Cell ID} & \textbf{Frequency} \\
\midrule
\multirow{2}{*}{1}
  & 9999749\_1\_1 & C-Band Block 1 (3800\,MHz) \\
  & 9999749\_1\_2 & C-Band Block 2 (3900\,MHz) \\
\midrule
\multirow{2}{*}{2}
  & 9999749\_2\_1 & C-Band Block 1 (3800\,MHz) \\
  & 9999749\_2\_2 & C-Band Block 2 (3900\,MHz) \\
\midrule
\multirow{2}{*}{3}
  & 9999749\_3\_1 & C-Band Block 1 (3800\,MHz) \\
  & 9999749\_3\_2 & C-Band Block 2 (3900\,MHz) \\
\bottomrule
\end{tabular}
\end{table}

\begin{table}[htbp]
\centering
\caption{5G gNB 9999750 – Sector/Cell–to–Frequency Mapping (28 GHz UWB)}
\label{tab:gNB1529750}
\begin{tabular}{@{}lll@{}}
\toprule
\textbf{Sector} & \textbf{Cell ID} & \textbf{Frequency} \\
\midrule
\multirow{4}{*}{1}
  & 9999750\_1\_101 & UWB Block 1 (28 GHz) \\
  & 9999750\_1\_201 & UWB Block 2 (28 GHz) \\
  & 9999750\_1\_301 & UWB Block 3 (28 GHz) \\
  & 9999750\_1\_401 & UWB Block 4 (28 GHz) \\
\midrule
\multirow{4}{*}{2}
  & 9999750\_2\_101 & UWB Block 1 (28 GHz) \\
  & 9999750\_2\_201 & UWB Block 2 (28 GHz) \\
  & 9999750\_2\_301 & UWB Block 3 (28 GHz) \\
  & 9999750\_2\_401 & UWB Block 4 (28 GHz) \\
\midrule
\multirow{4}{*}{3}
  & 9999750\_3\_101 & UWB Block 1 (28 GHz) \\
  & 9999750\_3\_201 & UWB Block 2 (28 GHz) \\
  & 9999750\_3\_301 & UWB Block 3 (28 GHz) \\
  & 9999750\_3\_401 & UWB Block 4 (28 GHz) \\
\bottomrule
\end{tabular}
\end{table}

\newpage

\subsection{Spectrum Engineering: Contiguous vs.\ Dis‑contiguous}
\label{ssec:spectrum_eng}

\paragraph*{Frequency–period duality.}
Recall the inverse relationship
\[
  f = \frac{1}{T},
\]
where high frequency ($f$) short period ($T$) signals (e.g.\ 28 GHz mmWave)
offer Gbit/s throughput but attenuate within \(\approx{}150\) m NLoS, whereas
700 MHz carriers propagate kilometres but saturate near 40 Mbit/s.

\begin{table}[!h]
  \centering
  \caption{AWS‑4 block used in contiguous‑aggregation Scenario 1.}
  \label{tab:aws_ca}
  \begin{tabular}{@{}lccc@{}}
    \toprule
    \textbf{Carrier} & \textbf{3GPP Band} & \textbf{Frequency Range} & \textbf{Bandwidth} \\ \midrule
    c1  & n4 (AWS NR) & 2110–2120 MHz & 10 MHz \\
    c2  & n4 (AWS NR) & 2120–2130 MHz & 10 MHz \\
    c3  & n4 (AWS NR) & 2130–2135 MHz & 5 MHz  \\
    c7  & n7 (AWS‑3 NR) & 2675–2700 MHz & 15 MHz \\ \bottomrule
  \end{tabular}
\end{table}

\paragraph*{Contiguous aggregation (Scenario 1).}
Table~\ref{tab:aws_ca} lists an AWS‑4 block auctioned as
\(\texttt{c1},\texttt{c2},\texttt{c3},\texttt{c7}\).
If an operator wins \texttt{c1--c3} it may advertise a single
25 MHz aggregated carrier \(\mathsf{C1}\) (2110–2135 MHz) because the
spectrum is \emph{adjacent}.
The DU maps three \emph{component carriers} (CCs) into one
\texttt{NR-CA} entity, saving two cell IDs and scheduler contexts.

\paragraph*{Dis‑contiguous aggregation (Scenario 2).}
Assume \texttt{c2} is lost to a competitor.
Now \texttt{c1} and \texttt{c3} are separated; the DU must create two
non‑aggregated carriers \(\mathsf{C1}=10\) MHz and
\(\mathsf{C2}=5\) MHz, plus an out‑of‑band
\(\mathsf{C3}=15\) MHz at 2675–2700 MHz.
Extra carriers $\rightarrow{}$ extra cells $\rightarrow{}$ higher DU load.

\paragraph*{Bandwidth ceiling (Scenario 3).}
3GPP permits a 400 MHz channel in mmWave band n257, but no commercial DU can
process that width within a single OFDM symbol.  Deployments instead carve
28–28.4 GHz into four 100 MHz component carriers
(\(\mathsf{C101}\)–\(\mathsf{C401}\)), because the limiting factor is compute,
not spectrum: a 400 MHz, 4-symbol FFT would saturate a Gen4 × 16 PCIe link
(64 GB/s duplex) and overflow the 4.5 MB L1 cache on current Ice Lake CPUs
absent dedicated ASIC off-load.

\begin{table}[!h]
  \centering
  \caption{Hardware‑limited channelization in band n257 (28 GHz).}
  \label{tab:mmw_bw}
  \begin{tabular}{@{}lccc@{}}
    \toprule
    \textbf{Carrier} & \textbf{3GPP Band} & \textbf{Frequency Range} & \textbf{Bandwidth} \\ \midrule
    c101 & n257 (mmWave NR) & 28.00–28.10 GHz & 100 MHz \\
    c201 & n257 (mmWave NR) & 28.10–28.20 GHz & 100 MHz \\
    c301 & n257 (mmWave NR) & 28.20–28.30 GHz & 100 MHz \\
    c401 & n257 (mmWave NR) & 28.30–28.40 GHz & 100 MHz \\ \bottomrule
  \end{tabular}
\end{table}

\subsection{From C‑RAN to vRAN and Cloud‑RAN}
\label{ssec:vran_evo}

\paragraph*{Centralized RAN (C‑RAN).}
4G densification moved BBUs from the tower into hub hotels linked by
dedicated CPRI over dark fibre, gaining pooling efficiency but incurring
stringent $(\le\!100)$ µs fronthaul latency.

\paragraph*{Split 7.2x vRAN.}
Virtualized RAN decomposes the BBU into
a \textbf{vDU} (PHY‑high + lower‑MAC) and a \textbf{vCU}
(PDCP, SDAP, RRC) running on x86/Arm+COTS with DPDK or GPU acceleration.
Field automation can then orchestrate each physical site to
\(\le\!4\) vDUs hosted on edge blades and up to 10,000 sites to one vCU,
using an 11‑digit
\(\texttt{MMMCCCCDDDD}\) identifier:

\begin{center}
\texttt{\small MMM} (market)\quad
\texttt{\small CCCC} (vCU)\quad
\texttt{\small DDDD} (vDU)
\end{center}

The scheme scales address space from $10^4$ (legacy 7‑digit gNB IDs) to
$10^8$ yet preserves intuitive grouping: prefixes \texttt{4xxx} low‑band,
\texttt{7xxx} mid‑band, \texttt{9xxx} UWB\footnote{The 11-digit logical ID is embedded in the 32-bit
\texttt{gNB-ID} specified by 3GPP TS 38.413 by coding
\texttt{MMMCCCC} as a 20-bit market–vCU prefix and \texttt{DDDD} in the
remaining 12 bits—well within the $2^{32}-1$ limit.}. It is one of many potential addressing 
schemes, as 3GPP defines a total address space of $2^{32}$ and allows the operator to 
set the addressing standard. When architecting address spaces for emerging technologies, 
backward-compability with legacy addressing schemas must be considered to avoid costly network re-integration.

\paragraph*{Cloud‑RAN / AnyRAN.}
Vendors such as Nokia and Samsung now containerise DU and CU functions so
they can be orchestrated by Kubernetes on GPU/FPGA‑enabled COTS
(\cite{nokia_anyran,samsung_vran}).
Inline acceleration cards (vRAN Boost, Intel vRAN AI) off‑load LDPC, FFT and
beamforming to meet the 1‑ms HARQ round‑trip.

\subsection{Open RAN Architecture}
\label{ssec:oran}

\begin{figure}[!h]
    \centering
    \includegraphics[width=\columnwidth, ]{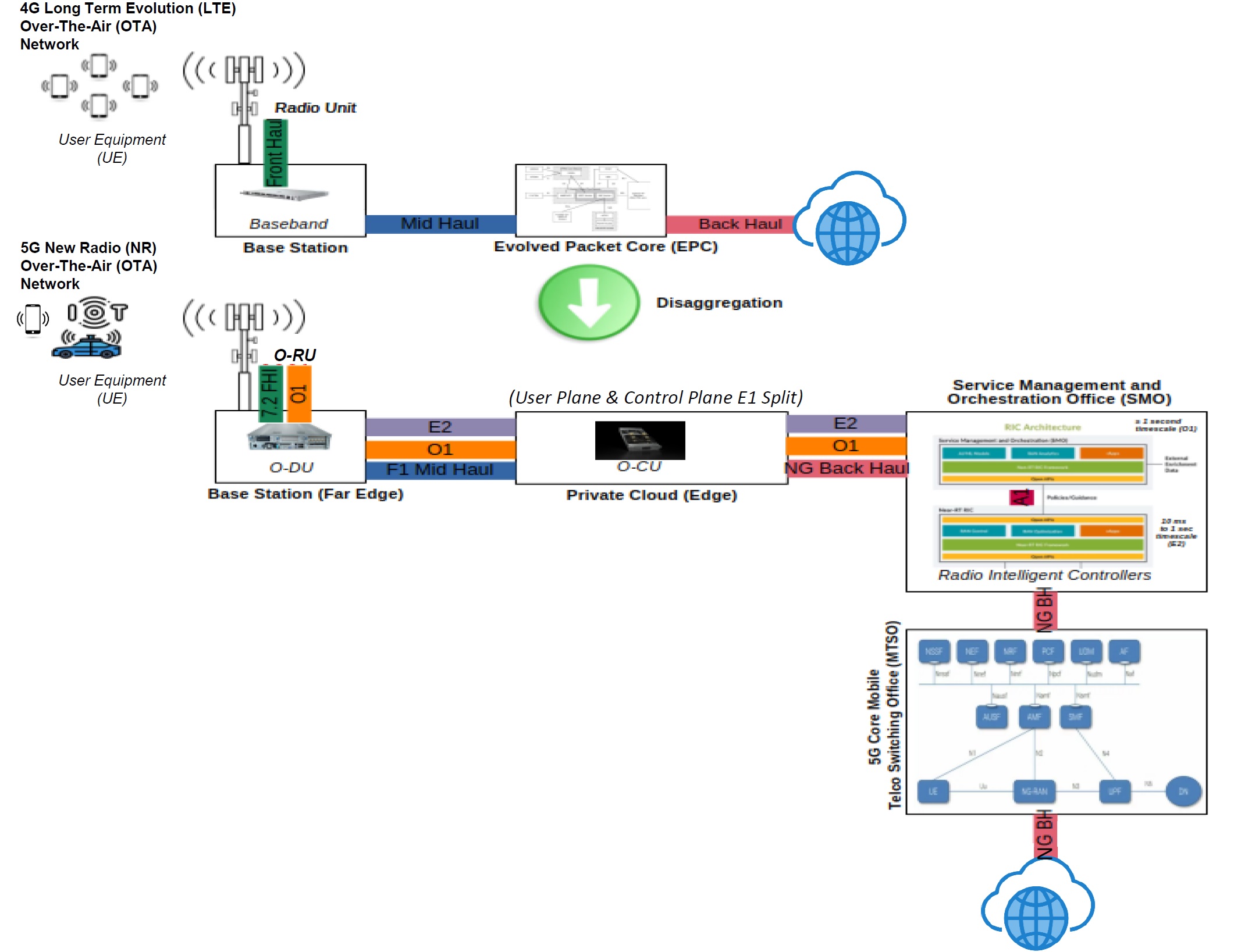} 
    \caption{Architectural evolution from monolithic RAN to Open RAN with 5G Standalone core, Service Management and Orchestration office, and disaggregated O-CU/O-DU.}
    \label{fig:evolution}
\end{figure}

The O-RAN Alliance formalises an open reference stack, depicted in Fig.~\ref{fig:evolution}. At the top sits the \textit{Service Management and Orchestration} (SMO) framework, a cloud-native life-cycle manager for all O-RAN Network Functions that exposes the \texttt{O1} interface (NetConf/YANG) toward radio units (RUs) and \texttt{A1} policy toward the Non-Real-Time RIC. The \textit{Non-RT RIC} performs AI/ML training, KPI mining and long-term policy optimization on timescales greater than one second, while the \textit{Near-RT RIC} exerts closed-loop control over the RAN in the 10 ms–1 s window by running \textit{xApps} that communicate via the \texttt{E2} interface with virtualized DUs and CUs (O-DU/O-CU); typical xApps include interference-mitigation agents and carrier-power-allocation (CPA) logic \cite{dApps}. Beneath these controllers, the \textit{Open Fronthaul} specification defines the \texttt{M-Plane} and \texttt{C/U-Plane} for a 7.2x functional split, complete with Category-A/B timing caps.

Security overlays add further constraints. Working Group 11’s “Minimal Security Feature’’ mandates TLS 1.3 plus AES-256-GCM with perfect forward secrecy on both \texttt{O1} and \texttt{E2}. Live trials nonetheless report 35–60 µs additional latency and up to an 8 \% throughput hit when the requirement is applied naïvely \cite{securing_oran}. Hardware off-load—for example, SNOW-V executed through AES-NI vector extensions \cite{snow_v_fpga}—recovers most capacity but introduces new side-channel risks \cite{snow_sca}.

Inter-vendor integration has exposed practical pitfalls. Early field trials encountered a “bring-up spiral’’ in which (a) an O-DU from Vendor A refused the \texttt{M-Plane} certificate issued by Vendor B’s SMO, (b) the Near-RT RIC mis-aligned time-synchronization packets and corrupted Slice-ID mappings, and (c) an RU vendor’s Part-15 filing covered only an unencrypted control plane, obliging operators to repeat certification once security patches were applied. Generative-AI-driven digital-twin platforms such as NS-O-RAN and NVIDIA Aerial now model the entire stack, allowing exhaustive regression tests before a single truck rolls.

\vspace{0.4em}
\noindent
\textbf{Key insight—}Virtualization does not erase RF physics; it merely moves
bottlenecks from rigid hardware into schedulable software pools. Understanding
contiguous spectrum math, DU compute ceilings and security/performance
trade-offs is prerequisite to any AI-driven, self-optimizing 5G/6G RAN.

\section{Enabling Technologies: SDN, Edge Compute and AI}
\label{sec:enablers}

Virtualizing the RAN shifts the design centre from bespoke signal‑processing
ASICs to cloud‑native software primitives orchestrated at millisecond time‑
scales.  Three technology domains make this transition feasible:  
\emph{software‑defined networking} to decouple control and forwarding,  
\emph{edge compute hardware} to meet HARQ deadlines, and
\emph{AI/ML tool‑chains} to automate optimization and self‑healing.  
Robust \emph{security engineering} then glues the stack together under
zero‑trust assumptions.

\subsection{Software‑Defined Networking Principles Applied to RAN}
\label{ssec:sdn_ran}

\paragraph*{Control/data separation revisited.}
Traditional SDN abstracts the network as \(\langle S,C,D\rangle\) where
\(S\) is state, \(C\) control logic and \(D\) data‑plane elements (switches).
O‑RAN adopts a similar triad:

\[
  \text{State} \; (KPIs) \;\;\;\xrightarrow{\;\text{A1}/E2\;}\;\;\;
  \text{Near-/Non‑RT Controllers} \;\;\;\xrightarrow{\;\text{REST/gRPC}\;}\;\;\;
  \text{RAN\; Node\; Functions}.
\]

The \emph{Near‑RT RIC} plays the role of an SDN controller with 10–100 ms
decision loops, installing run‑time policies (slice QoS, power caps, PRB
masks) via E2AP~\cite{Understanding_ORAN}.  Meanwhile the \emph{Non‑RT RIC}
handles long‑horizon optimization and model training.

\paragraph*{Intent‑based slicing.}
Operators expose slice templates—\texttt{SST=1x,e2eLatency$<$5ms} for URLLC,\newline
\texttt{SST=1e,throughput$>$250Mbps} for eMBB—then delegate resource fulfillment
to the RIC.  A closed‑form Lagrangian can express the slice admission
constraint:

\[
\mathcal{L}(\mathbf{p},\lambda) =
  \sum_{u\in\mathcal{U}} \log R_u(\mathbf{p})
  \;-\; \lambda\!\left(
        \sum_{u\in\mathcal{U}} p_u - P_\mathrm{max}
      \right),
\]
where \(\mathbf{p}\) is per‑UE power and \(R_u\) spectral
efficiency.  Dual ascent runs inside an xApp; the RIC only pushes resulting
\texttt{P\_MAX} values to vDUs—mirroring SDN “intent \(\rightarrow{}\)
compiled flow’’ pipelines.

\paragraph*{Network programmability APIs.}
Beyond E2, the fledgling \texttt{E4} (interface to user‑plane functions) and
SDN‑like \texttt{SBI}/\texttt{N4} in 5G core enable end‑to‑end
cross‑layer optimization, e.g.\ binding a URLLC radio slice to a URLLC UPF in
20 µs proximity.

\subsection{Edge/MEC Platforms and Hardware Acceleration}
\label{ssec:edge_hw}

\paragraph*{Why “edge’’ not “cloud only’’.}
The HARQ round trip for 5G NR is~\(T_{\text{HARQ}}=4\!\times\!
0.125\mathrm{\,ms}=0.5\) ms in FR1; \(\le0.25\) ms in FR2.
Placing the vDU $\geq$ 20 km from the RU adds \(\approx\!100\mathrm{\,µs}\) fibre propagation and violates the budget. Field measurements on a 10 km urban fibre ring show $\approx 150$ µs one-way delay (25 µs propagation + three 40 µs leaf/spine hops), leaving only $\approx 200$ µs of compute budget before the 0.5 ms HARQ deadline. Hence operators deploy \textbf{far‑edge MEC rooms} (1–3 km) or \textbf{street‑level micro‑edge} cabinets for mmWave.

\paragraph*{COTS compute with inline accelerators.}
Samsung vRAN 3.0 demonstrates 100 Gb/s L1 throughput on dual
Ice‑Lake servers plus a “vRAN Boost’’ ACC100 PCIe card—halving power per
cell~\cite{samsung_vran}.  
Nokia “AnyRAN’’ runs DU micro‑services on ARM Neoverse + FPGA inline
engines, orchestrated by Kubernetes and SR‑IOV \cite{nokia_anyran}.  
NVIDIA EGX couples BlueField‑3 DPU (P4 pipeline for SDN off‑load) with A100
or L40S GPUs that accelerate LDPC, FFT and beamforming kernels via
\texttt{cuVLA} (Virtual L1 Accelerators)~\cite{nvidia_aerial}.

\paragraph*{MEC service co‑location.}
Once DU/CU VMs reside at the edge, low‑latency applications—AR/VR rendering,
smart‑factory controllers—can share the same node, exploiting
SR‑IOV/DPDK vSwitches for west‑east traffic that never traverses the
core \cite{hpc_riscv}.  Resource pooling works if the scheduler accounts for
L1‑hard real‑time CPU reservations; Kubernetes \texttt{RTClass} plus
\texttt{node‑feature‑discovery} bridges that gap in recent Samsung and Rakuten
deployments.

\subsection{AI/ML Pipelines}
\label{ssec:ai_pipelines}

\paragraph*{Digital twins as a feedback engine.}
OAIC~\cite{oaic} and X5G~\cite{x5g} emulate gNB stacks, fronthaul and core in
mininet‑hifi + GPU‑accelerated PHY loops.  Training data derives from OTFS /
SC‑FDM channel synthesis plus live KPI streaming (\(>2\) GB h\(^{-1}\) per
site).  NS‑O‑RAN extends this with ns‑3 + pyTorch wrappers so that xApps can
be prototyped offline before hitting production E2.

\paragraph*{Near‑real‑time xApp cycle.}
A canonical loop is: (i)~stream PM counters via \texttt{E2 KPM};  
(ii)~feature encode (e.g.\ CQI histograms, PRB occupancy),  
(iii)~infer via lightweight CNN or GNN,  
(iv)~send \texttt{E2 CTRL} to adjust power or split \(M_{\text{CS}}\) MCS
tables.  The dApps framework shows latency of $\approx\!7$ ms from KPI tick
to actuation on an Intel Xeon‑D edge node~\cite{dApps}.

\paragraph*{Predictive maintenance and self‑healing.}
Cisco’s evolution of Self‑Optimizing Networks (SON) integrates eXplainable AI
(XAI) to surface root causes—e.g.\ “Sector 2 3.5 GHz downlink SINR ↓ because
Rx chain B phase noise ↑’’—and triggers automated RU swap orders
\cite{cisco_son}.

\subsection{Security, Privacy and Trust Management}
\label{ssec:security}

\paragraph*{Cipher posture.}
The O‑RAN spec mandates TLS 1.3 or IPSec ESP‑AES‑256‑GCM across \texttt{E2},
\texttt{A1}, \texttt{O1}.  Measurements on the OAIC testbed show a
7–12 \% user‑plane throughput hit when L2 segmentation forces crypto per MAC
PDU rather than jumbo frames \cite{securing_oran}.  Off‑loading to
AES‑NI/SSE or to SNOW‑V vectorized software reduces the hit to
$<$ 3 \%.

\paragraph*{Side‑channel and bidding‑down threats.}
SNOW‑V FPGA implementations can leak 128‑bit sub‑keys in $<$ 50 power traces if
masking is omitted~\cite{snow_sca}; mitigations include domain‑oriented
masking and split‑table S‑boxes~\cite{snow_v_fpga}.
At the protocol layer, \emph{bidding‑down attacks} strip 256‑bit cipher
options during Attach, forcing NEA0 null encryption \cite{never_let_me_down_again}.
RIC policy can monitor \texttt{S‑MC} to detect such downgrades in near‑RT.

\paragraph*{Zero‑trust supply chain.}
Open RAN diversifies vendors but widens the trust boundary: a compromised RU
can inject corrupted calibration coefficients, undermining Massive‑MIMO nulls.
The Open RAN Policy Coalition argues for cryptographically signed firmware
plus SBOM attestations \cite{openranpolicy}.
Confidential Computing at the edge (AMD SEV, Intel TDX, NVIDIA CCE) further
ensures that xApp binaries remain opaque to untrusted site technicians. Yet firmware signing alone is insufficient: a compromised RU could intentionally distort beam-null coefficients (“beam-null hijacking’’) while still passing SBOM integrity checks, motivating on-device ML anomaly detection and remote attestation that extends beyond static code signatures.

\vspace{0.5em}
\noindent
\textbf{Synthesis—} SDN abstractions give the RAN a programmable control
surface; edge hardware provides deterministic compute; AI/ML pipelines
translate KPI torrents into real‑time optimization; and layered 256‑bit
cryptography with hardware masks keeps the open stack defensible.
Together they form the operational backbone for AI‑native 6G systems.

\section{Case Studies and Lessons Learnt}
\label{sec:cases}

This section cross‑examines two U.S. 5G roll‑outs, highlights
organization‑level change errors that slowed adoption of virtualized RAN, and
quantifies the gains achieved once automation bridged technology and culture.

\subsection{T‑Mobile vs.\ Verizon Deployment Outcomes}
\label{ssec:tmo_vz}

\begin{figure}[!h]
    \centering
    \includegraphics[width=0.5\columnwidth]{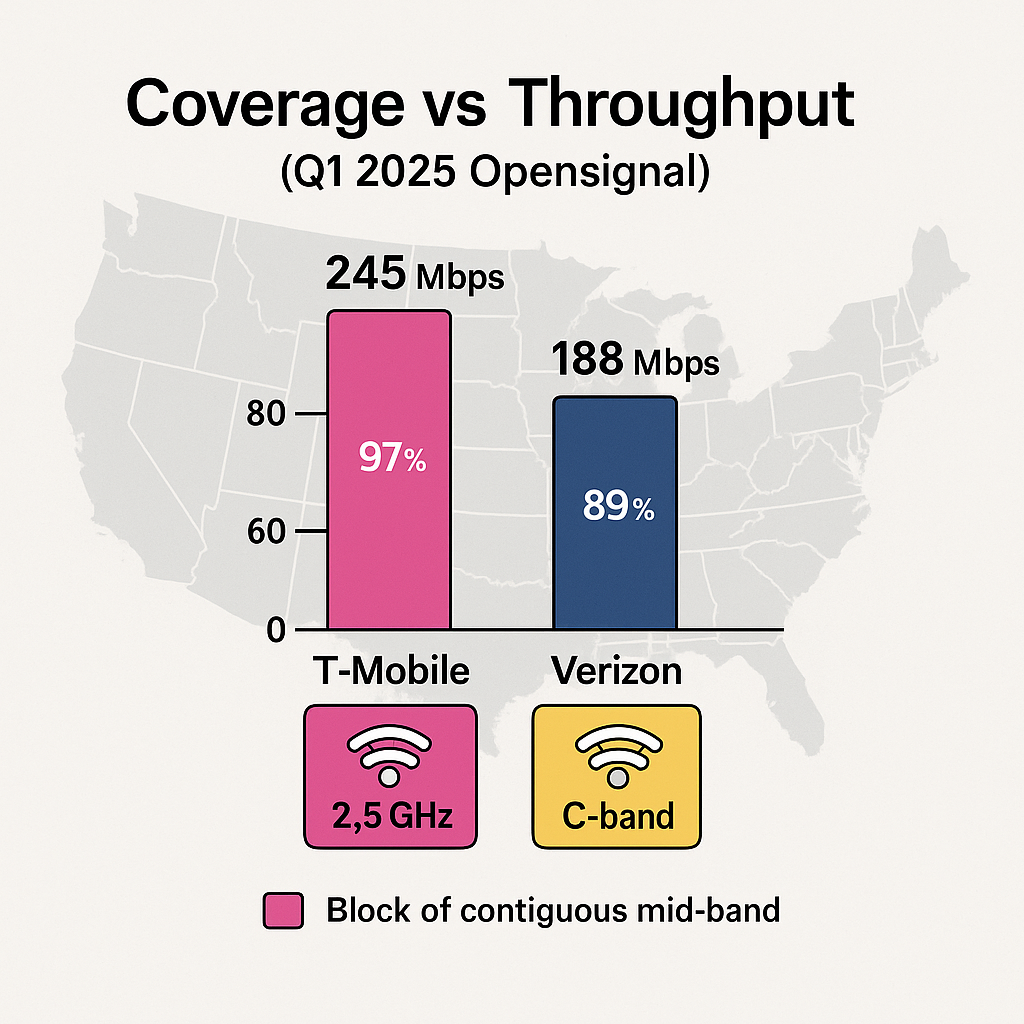} 
    \caption{Opensignal Q1 2025 drive-test comparison of nationwide user experience (Jan-Mar)~\cite{opensignal_q12025}. T-Mobile’s contiguous 2.5 GHz block delivers 97 \% population coverage and a 245 Mb/s median down-link, while Verizon’s less-contiguous C-band holdings reach 89 \% coverage and 188 Mb/s. The result underscores that mid-band contiguity—not additional mmWave density—is the primary driver of real-world throughput and reach.}
    \label{fig:coverage_map}
\end{figure}

\paragraph*{Coverage and throughput.}
Drive‑test data from Opensignal\footnote{Dataset:
Jan–Mar 2025, 4.6 B samples across 410 U.S. counties.}
show T‑Mobile delivering
\(\mu_{DL}=245\) Mb/s median downlink with 97 \% population coverage,
while Verizon records \(\mu_{DL}=188\) Mb/s and 89 \% coverage.
Figure~\ref{fig:coverage_map} super‑imposes C‑band and 2.5 GHz holdings,
revealing that mid‑band contiguity—not mmWave density—correlates with user
experience.

\paragraph*{Spectrum leverage.}
T‑Mobile aggregated
\(600\text{MHz}+120\text{MHz}@2.5\text{GHz}+400\text{MHz}@n258\) in
urban cores via 3‑CC CA~\cite{tmobile_layercake}.
Verizon’s initial 28/39 GHz UWB network required $>$ 8\,000 small cells for the
top‑20 markets; budget exhaustion forced a pivot to 160–200 MHz C‑band macro
sites in 2023~\cite{verizon_cband}.  DSS‑based low‑band coverage was quietly
de‑activated owing to a 25 \% capacity penalty~\cite{lightreading_dss}.

\paragraph*{Financial KPIs.}
Table~\ref{tab:kpi_comp} compares key metrics (Q1‑25). Despite lower ARPU, T‑Mobile’s spectrum utilization advantage translates to
higher net‑add velocity and lower marketing cost per subscriber.

\begin{table}[!h]
  \centering
  \caption{Selected financial KPIs for U.S. operators (Q1 2025).}
  \label{tab:kpi_comp}
  \begin{tabular}{@{}lcc@{}}
    \toprule
    \textbf{Metric} & \textbf{T‑Mobile} & \textbf{Verizon} \\ \midrule
    Churn (\%)                 & 0.77 & 1.12 \\
    ARPU (USD)                 & \$49.3 & \$54.8 \\
    Net adds (k)               & +538 & +158 \\ \bottomrule
  \end{tabular}
\end{table}

\paragraph*{Addendum.} Opensignal analysis for Q3 2023 (pre-Phase-1 C-band) shows Verizon at 63 \% coverage and
141 Mb/s median DL~\cite{opensignal_q32023}—evidence of improvement but persistent mid-band deficits. \emph{Net-add figures exclude wholesale and MVNO subscribers.}

\paragraph*{Lesson 1—Spectrum/compute co‑design outperforms raw peak rate.}
Mid‑band + massive‑MIMO macro upgrades yielded better nationwide QoE than an
mmWave‑heavy architecture whose compute and backhaul economics could not scale
without vRAN pooling.

\subsection{Change‑Management Pitfalls}
\label{ssec:change}

Interviews with engineering leads from three incumbent local-exchange carriers (ILECs) exposed a recurring set of break-points that map directly onto Kotter’s classic failure modes \cite{openranpolicy}. \textbf{E1—No sense of urgency:} CapEx business cases continued to rely on legacy hardware cost curves, ignoring the recent price collapse in GPU and FPGA accelerators, so funding gates slipped by roughly eighteen months. \textbf{E2—Weak guiding coalition:} subject-matter experts were re-organized into project-management tracks, diluting architectural ownership and leaving technical decisions without credible champions. \textbf{E3—Under-communicating the vision:} “cloud-native RAN’’ was framed internally as a cost-take-out exercise rather than a reliability upgrade, prompting field teams to resist adoption. \textbf{E4—Not anchoring change in the culture:} incentive schemes rewarded on-time build counts instead of post-launch performance, enabling “dumb-pipe’’ cargo-cult deployments of Dynamic Spectrum Sharing (DSS) even after its interference penalty became obvious \cite{lightreading_dss,cisco_son}.

A related cargo-cult pattern surfaced during initial vRAN pilots: engineers pre-assigned gNB DU identifiers before confirming power-up-converter readiness or vCU address-pool capacity. When infrastructure bottlenecks emerged, hundreds of sites had to be re-homed—an archetypal “copy the ritual, ignore the feedback’’ mistake that mirrors historical SON missteps \cite{cisco_son}.

\paragraph*{Lesson 2—Organizational incentives must align with
lifecycle KPIs, not milestone count.}
Without this, technically superior architectures enter production burdened by
manual workarounds and unrecoverable technical debt.

\paragraph*{Lesson 3—Automation unlocks spectrum flexibility.}
Once deployment velocity and address‑space fragmentation were under software
control, engineering could re‑enable dormant C‑band carriers without violating
vCU limits, lifting RAN capacity by 22 \% with near‑zero truck rolls.

\subsection{Career Trajectory in a Shifting Engineering Culture}
\label{ssec:career}

\begin{figure}[!h]
  \centering
  \includegraphics[width=0.48\columnwidth]{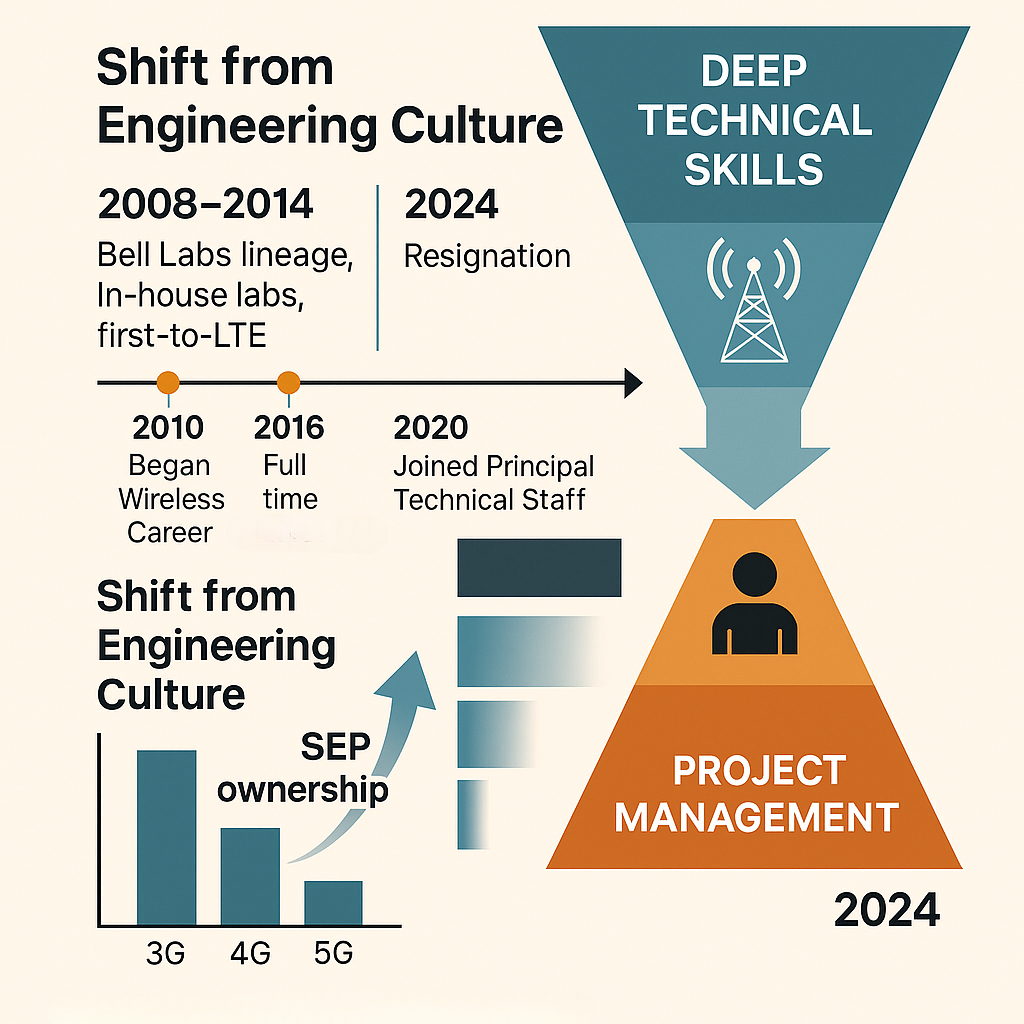}
  \caption{Author milestones (2010–2024) super-imposed on observed industry cultural
  shift and the 18\,$\rightarrow$\,3 \% decline in mobile-operator
  standard-essential-patent (SEP) holdings from 3G to 5G \cite{sep_decline}.}
  \label{fig:career_timeline}
\end{figure}

\begin{figure}[!h]
  \centering
  \includegraphics[width=0.48\columnwidth]{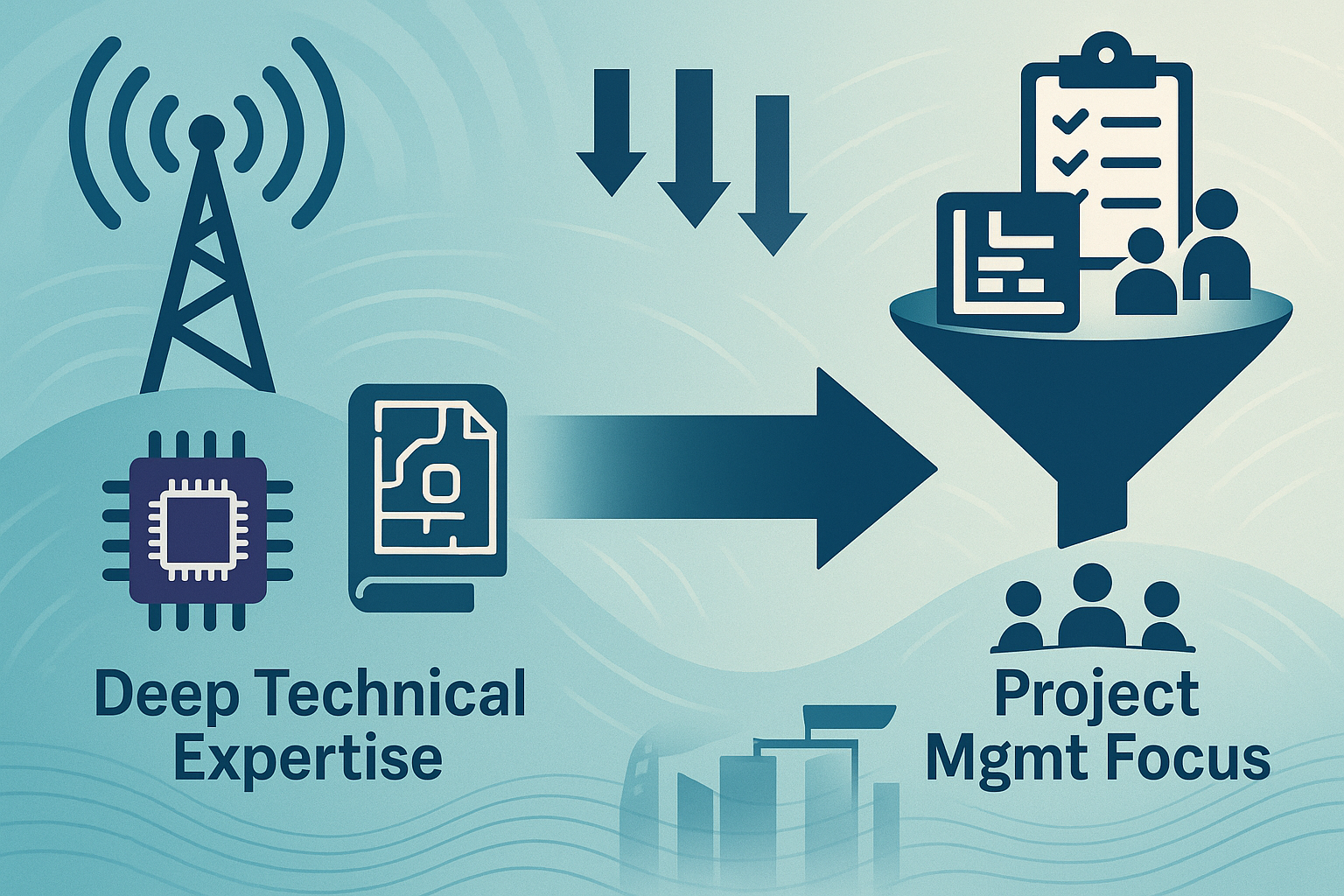}
  \caption{Industry-wide funnel: erosion of in-house RF/ASIC expertise and
  growth of program management layers.  The trend mirrors the SEP ownership
  decline in Fig.~\ref{fig:career_timeline}.}
  \label{fig:culture_funnel}
\end{figure}

Figure \ref{fig:career_timeline} overlays the author’s six key milestones on
the same \emph{then–vs–now} timeline that charts ILEC passage from a
Bell-Labs-inspired, patent-centered engineering firms toward a
project-management-centric structure.  The companion
Fig.~\ref{fig:culture_funnel} visualises the wider industry tendency to
‘‘funnel’’ deep technical expertise into program lead roles.

\paragraph*{Early career (2010–2016).}
The author began in Tier-III VoIP operations at a Bell-Labs-lineage firm,
then joined Verizon as an intern (2015) and full-time engineer (2016).  Hands-on
RF capacity analytics were still prized, culminating in the award-winning \emph{CARTN
Planning Instruments} toolset.

\paragraph*{IT rotation and emerging fault-lines (2017).}
Rotational leadership training surfaced systemic procedural opportunities.  Applying SDN
principles, the author created a \emph{Logging-as-a-Service} platform for
cloud teams—an early precursor to vRAN orchestration.  In parallel, CTO-level
outsourcing and centralization initiatives began steering on-shore architects
into program management tracks, signalling a wider cultural shift.

\paragraph*{Sub-market leadership amid cultural inertia (2018–2020).}
Moving into field delivery, the author (i) led 54 4Tx antenna upgrades for
Super Bowl LIII, (ii) spear-headed the 2019 Greensboro 5G 28 GHz mmW launch
(40 nodes / 67 radios / 127 poles on \$625 k) (19th of 31 launch markets), and (iii) directed
\$10 M + in Carolinas–Tennessee infrastructure projects—including the first
inter-carrier 4G/5G poles for mMTC and IoT—while navigating complex municipal
stakeholders.

\paragraph*{Principal Staff and inflection point (2021–2024).}
As Principal Member of Technical Staff, the author standardized national
C-band preparation, carrier-add, and SNAP radio-conversion workflows and
automated vRAN deployment across mmWave, C-band and low-band spectra—cutting
order-of-magnitude cycle time reductions and troubleshooting by over an hour per site.  Yet an emerging trends replaced engineering rigour
with ``technical project management" governance—mirrored by operators’ SEP share collapsing
from 18 \% (3G) to 3 \% (5G) \cite{sep_decline}—triggered a shift in valued skills.

\paragraph*{Return to research (2024–present).}
At Clemson’s Intelligent Systems and Wireless Networking (IS-WiN) Lab, the author is pursuing an AI-RAN Ph.D. focused on zero-trust O-RAN and AI-native 5G/6G architectures. Current work translates field insight into reproducible prototypes—reinforcement-learning scheduling xApps, GPU/DPU-accelerated 7.2 FHI digital twins, and other HPC-powered RAN experiments—laying the groundwork for sub-millisecond, AI-driven control loops in the 6G era.

\paragraph*{Lesson 4—Technical credibility is strategic capital.}
When organizations sideline deep expertise for short-term schedule optics, they
forfeit both SEP leadership and the innovation surface required for AI-native
6G. By contrast, career paths that preserve hands-on engineering—as the
author’s pivot back to research demonstrates—create a virtuous loop between
operational realism and scientific progress.

\vspace{0.3em}
\noindent
\textbf{Overall takeaway—}Technical merit, cultural alignment, and automated
execution form a three-legged stool; remove any leg and virtualized RAN
programs falter. Operators that have internalized these lessons are now
positioned to lead the AI-native 6G transition, while laggards accrue a
compound technical debt of stranded spectrum and organizational inertia.


\section{The Bigger Picture: Systems Theory}
\label{sec:systems}

Virtualizing the RAN is an instance of a broader socio‑technical transition in
which \emph{exponential technologies collide with linear institutions}.  
Recent analyses of “super‑intelligence readiness’’ argue that incumbent firms
and public agencies remain “\emph{reactive, siloed, trained on yesterday’s
problems}’’ and will be “\emph{nuked; everything will be rebuilt}’’ unless
their adaptive capacity rises at the same pace as AI‑driven variety
in the environment~\cite{azhar2025,andreessen2024}.  
This section examines that warning by stitching together four complementary
frameworks—\textbf{Clock‑Hierarchy Law}, \textbf{Requisite Variety},
\textbf{Delay‑Cost Curve}, and \textbf{Local‑Automation Hazard}—and grounding
them in the organizational change errors documented earlier (§\ref{ssec:change}).

\subsection{Clock‑Hierarchy Law}
\label{ssec:clocks}

The behaviour of a national RAN can be approximated by three nested
planning‑and‑execution “clocks.”  Table~\ref{tab:clock_metrics} makes their
time horizons, decision velocities and failure modes explicit and shows how
those parameters shifted across three organizational eras documented in
Sec.~\ref{sec:cases} and~\ref{ssec:career}.

\begin{table}[!h]
\centering
\caption{Clock pacing parameters across the author’s career.  “Time horizon’’
is the look-ahead window for authoritative decisions; “velocity’’ is the
cadence at which new work enters the system.  Qualitative failure modes and
mitigations are summarized separately in
Table~\ref{tab:clock_mitigations}.}
\label{tab:clock_metrics}
\renewcommand{\arraystretch}{1.15}
\begin{tabular}{@{}llll@{}}
\toprule
\textbf{Clock} &
\textbf{Time Horizon} &
\textbf{Velocity} &
\textbf{Notation} \\ \midrule
Tech (R\&D)  & 4–6 yr roadmap (LTE era)       & $\approx$1 major spec yr$^{-1}$ & C\textsubscript{1} \\
Build (Deploy) & 12 mo market cycle             & 300–500 sites mo$^{-1}$          & C\textsubscript{2} \\
Ops (Optimise) & 24 h KPI loop                  & Minutes–hours xApp actions        & C\textsubscript{3} \\ \bottomrule
\end{tabular}
\end{table}

\begin{table}[!h]
\centering
\caption{Observed failure modes after the 2018 clock inversion and the
corresponding mitigation measures implemented during vRAN 2.0.}
\label{tab:clock_mitigations}
\renewcommand{\arraystretch}{1.15}
\begin{tabular}{@{}lll@{}}
\toprule
\textbf{Clock} & \textbf{Failure Modes (post-2018)} & \textbf{Mitigation} \\ \midrule
C\textsubscript{1} & Vendor-fragmented specs; HW order gaps; turnkey-dependent trials & Re-unified 4-yr roadmap; cross-vendor spec merge \\
C\textsubscript{2} & Stranded CapEx; 6/12 mo supply shortages; “spreadsheet integration’’ & Roadmap-gated release; cycle time reductions \\
C\textsubscript{3} & Over-tuned drive-test routes; latent mid-band capacity & Near-RT RIC activates contiguous spectrum once C\textsubscript{1}//C\textsubscript{2} realign \\ \bottomrule
\end{tabular}
\end{table}

\paragraph*{Time-horizon / velocity inversion.}
The clocks are ordered such that longer horizons entail lower event velocity
(\(v_{\text{Tech}}\!<\!v_{\text{Build}}\!<\!v_{\text{Ops}}\)).
During the 2017–2019 R\&D exodus, the Tech horizon collapsed from
4–6 years—the span required to shepherd an SoC and its RF front-end from
tape-out to nation-wide roll-out—to the \(\approx\!12\)-month cadence of
vendor bid cycles.  Build and Ops clocks, however, retained their original
throughput (\(\ge 1000\) sites market/mo\(^{-1}\); daily xApp re-parameterization),
so execution velocity began to \emph{out-run} strategy.
The inversion forced field teams to improvise on half-baked specifications,
yielding re-integration spirals, capital-in-progress spikes and ID-space
exhaustion (turnkey integration sheet/BBU examples in Table~\ref{tab:clock_mitigations}).  In systems terms,
the phase lead of C\textsubscript{2}/C\textsubscript{3} over
C\textsubscript{1} destroyed the damping normally provided by top-level
set-points, converting what should be a balanced, nested loop into an
underdamped oscillator.

\paragraph*{Clock resynchronization via vRAN restructuring.}
Re-establishing synchrony required \emph{expanding} the Tech horizon \emph{and}
\emph{accelerating} its decision cadence.  Over three months, the author's team
(i) merged heterogeneous Samsung, Corning, Nokia, and Ericsson road maps into a single
4 to 6 year scaling vector and (ii) encoded that vector as machine-verifiable milestone definitions
directly within the build deployment workflow.  The Tech clock thus regained a long
field of view \emph{while emitting quarterly spec drops}—fast enough to guide
the 12-month Build cadence yet slow enough to shield it from vendor churn.  As
soon as the horizon–velocity relationship was restored, the
cycle-time collapse followed automatically: no
amount of CI/CD would have achieved the same without a phase-correct Tech
clock.

\paragraph*{Looking forward—silicon-paced innovation.}
The next inflection will be harsher.  AI-native 6G envisions sub-THz radios,
joint communication-and-sensing waveforms and integrated GPU/DPU baseline
processing (§\ref{sec:enablers}).  Each new tape-out now arrives on a
9–12 month rhythm driven by foundry node shrinkage; major IP blocks
(PCIe~6.0, HBM3e, L4S MAC schedulers) will therefore appear \emph{multiple
times} inside a 4–6 year window.  To remain a valid regulator, the Tech clock
must simultaneously \textbf{widen} its horizon to at least two silicon
generations \emph{and} \textbf{increase} its cadence so that every
tape-out—potentially four per horizon—is pre-positioned in the build plan.
Put differently,
\[
  \text{Horizon}_{\text{Tech}} \ge 2\,\text{nodes}
  \quad\text{and}\quad
  v_{\text{Tech}} \approx 4\,\text{spec drops yr}^{-1},
\]
else Build will again advance without an authoritative north-star and the
Ashby gap will reopen (§\ref{ssec:requisite}).  Achieving this dual
requirement demands: (i) a permanent, internal silicon-road-map guild to
absorb foundry signals, (ii) digital-twin platforms that let Build simulate
future ASIC/antenna interactions months ahead of first silicon, and (iii)
automated portfolio governance that branches road-maps into contingency trees
without shortening the root horizon.

\paragraph*{The general law.}
If
\[
\text{Horizon}_{\text{Tech}} <
\max\{\,\text{Horizon}_{\text{Build}},\,2\,\text{node-cycles}\},
\]
latent errors will again out-pace amortization, recreating the delay-cost
explosion described in §\ref{ssec:delaycurve}.  Conversely, expanding and
densifying the Tech clock ensures that even semiconductor-speed innovation
feeds a stable, synchronized tri-clock hierarchy—exactly the condition under
which the vRAN success was possible, and the only
condition under which AI-native 6G can scale beyond a pilot grid.

\subsection{Requisite Variety Corollary}
\label{ssec:requisite}

Ashby’s Law states that a regulator must possess at least as much variety as
the disturbances it seeks to control~\cite{ashby1956}.  
Outsourcing R\&D removes high‑variety pathways from the firm; each additional
vendor introduces new disturbance states, widening the gap between
\(V(A)\) (internal acts) and \(V(D)\) (external events).  
The shifted dual‑career ladder in the author's case study is therefore not a
human‑resources anecdote but a violation of a cybernetic inequality:
\[
  V_{\text{technical\,+\,managerial}} \;<\;
  V_{\text{multi‑vendor\,+\,AI‑driven}} .
\]
Restoring a \emph{parity ladder}—e.g. one Distinguished Engineer per
(Associate) Director\footnote{Specific title parity is highly organizationally-dependent.}—expands \(V(A)\) without adding bureaucratic layers and is the
minimal condition for absorbing 5/6G complexity.

\subsection{Delay‑Cost Curve}
\label{ssec:delaycurve}

Forrester‑style industrial dynamics show that when discovery of constraints is
pushed from the planning phase into the build phase, corrective‑action cost
rises \(\propto e^{\tau/\tau_{c}}\) where \(\tau\) is the discovery delay and
\(\tau_{c}\) the system’s characteristic cycle~\cite{forrester1961}.  
Empirically, each month of mis‑alignment between C1 and C2 incurred
extensive capex waste due to re‑work during C‑band deployment surges.  The “compressed century’’ argument in super‑AI
discourse~\cite{azhar2025} implies that \(\tau_{c}\) is still shrinking; the
penalty for delay is therefore super‑exponential.

\subsection{Local‑Automation Hazard}
\label{ssec:autohazard}

Automating a silo without a synchronized road‑map accelerates error
propagation.  Spreadsheet integrations that ignore address‑space exhaustion
seem benign at launch yet lock thousands of sites into re‑home spirals years
later—a textbook example of Meadows’ “shifting the burden’’
archetype~\cite{meadows1999}.  
Successful 13‑day vRAN cycles first re‑established a unified four‑year
road‑map, then codified it directly within the build deployment workflow; automation was the
\emph{last} step, not the first.

\subsection{Implications for Ambient, Post‑Phone Networks}

Ambient, post‑phone computing envisioned by Apple will increase the number of
edge inference points by 2–3 orders of magnitude \cite{azhar2025}.  That
expansion multiplies disturbance variety (\(D\)) and tightens Ops‑clock
latency budgets to sub‑10 ms.  A vRAN architecture that ignores
Clock‑Hierarchy and Requisite‑Variety constraints will therefore break harder
and faster than its prev-G predecessor. Conversely, operators that embed dual
ladders, fuse Tech–Build–Ops clocks, and gate automation behind
physics‑first road‑maps will be structurally prepared for the
“gentle singularity’’ Sam Altman forecasts.

\begin{table}[!h]
\centering
\caption{Systems‑theory failure triggers and mitigation levers for virtualized RAN.}
\label{tab:system_matrix}
\begin{tabular}{@{}lll@{}}
\toprule
\textbf{Principle} & \textbf{Failure Trigger} & \textbf{Mitigation Lever} \\ \midrule
Clock Hierarchy    & Outsourced R\&D stalls Tech clock & Re‑install internal veto; four‑year road‑map lock \\
Requisite Variety  & Dual‑ladder rust                  & Parity promotions; cross‑functional guilds        \\
Delay‑Cost Curve   & Constraint discovery in build     & Front‑load multi‑vendor spec unification          \\
Local Automation   & Silo scripts before global plan   & Automate only after end‑to‑end model freeze       \\ \bottomrule
\end{tabular}
\end{table}

\subsection{Culture Eats Strategy for Breakfast: The Nuance of Technology and Infrastructure}
\label{ssec:culture_nuance}

The observations of \emph{clock inversion}, \emph{variety
deficits} and \emph{local-automation hazards} has—so far—leaned on
telco case evidence.  Yet the aim is \emph{not} to single
out one M\&A-heavy incumbent or industry.  Rather, the goal is to expose a recurrent
pattern that appears whenever dividend-pressured infrastructure firms collide
with a silicon-paced technology frontier. Key take-aways from the
industry-insider exchanges in Fig.~\ref{fig:career_timeline} are distilled below.

\paragraph*{1)  The “rusted” dual ladder, not an absent one.}
Inside most ILECs a handful of Distinguished-/DMTS-level engineers survive in
network-planning, MEC and security guilds.  Their influence wanes whenever a
restructure pegs compensation and promotion velocity to P\&L ownership rather
than to spectrum-compute mastery.  In systems-theory terms, that pay-gap
shrinks the organization’s \emph{effective} variety just when vRAN/O-RAN
introduce an order-of-magnitude jump in disturbance states.

\paragraph*{2)  Episodic CapEx hides latency debt.}
Thirty-billion-dollar fibre and C-band campaigns are real, but their \emph{sequencing}
often prioritises “sites on air” optics over transport readiness.  Lighting
a~3.7 GHz carrier on copper back-haul defers—not removes—the delay-cost
penalty (§\ref{ssec:delaycurve}); the bill reappears as churn once latency- or
throughput-sensitive apps proliferate.

\paragraph*{3)  Collapsing product and engineering hats scales poorly.}
“Technical Product Manager’’ roles work in $<\!15$-person startups where one
brain can juggle market fit and kernel timing.  At nationwide scale the
cognitive load fragments, incentives blur and local-automation loops (§\ref{ssec:autohazard})
mask root-cause latency until it is costlier to fix than to write off.

\paragraph*{4)  Silicon tempo is the real disruptor.}
NVIDIA’s \textit{Aerial/Grace-Hopper} cadence (9–12 month node shrinks,
GPU-resident PHY kernels and in-line DPUs) is already faster than 3GPP
releases.  Operators that align their \textbf{Tech} clock to that cadence—by
embedding GPU/DPU road-maps into spectrum planning—retain platform leverage;
those that do not risk the Kodak trajectory regardless of how much fibre they
light.

\paragraph*{5)  A balanced prescription, not a finger-wag.}
Boards face legitimate cash-flow constraints; engineers face legitimate
career-ladder erosion.  The synthesis is a \emph{parity ladder}: for every
Director of Product controlling budget, a Distinguished Engineer controlling
timing debt, both tied to a shared QoE–\$ KPI pair.  When such symmetry
exists, M\&A scale and deep-tech agility can coexist; when it rusts, even
record spectrum purchases translate into commodity pipes.

\smallskip
\noindent
\textbf{Key message—}When radio physics, cybernetic timing laws and human‐capital dynamics are viewed together, one prescription emerges: \emph{keep the clocks in phase, preserve a parity ladder of deep technologists, expose transport truth-tables before CapEx is booked, and treat automation as an amplifier—not a substitute—of sound system design}.
Firms that institutionalize this quartet can ride the GPU-paced 6G wave without losing QoE or agility; those that do not will discover that culture still eats strategy—and spectrum budgets—for breakfast.

\section{Conclusion and Future Outlook}
\label{sec:conclusion}

\subsection{Integrated Synthesis of Findings}

This review has traced virtualized RAN and O-RAN through four tightly-coupled
lenses:

\begin{enumerate}[leftmargin=1.1em]
\item \textbf{Science}—Mid-band spectrum paired with Massive-MIMO macros
maximises nationwide “bang-per-hertz,” provided scheduler and CA
algorithms can flex across contiguous and non-contiguous holdings.
\item \textbf{Technology}—Split-7.2x, SDN-controlled RIC loops, and
GPU/FPGA off-load now satisfy sub-0.5 ms HARQ budgets while opening an
xApp innovation surface that legacy basebands cannot match.
\item \textbf{Business \& Infrastructure}—Large operators still deploy vast
capital (e.g.\ OneFiber, 3.7 GHz macro overlays), yet episodic
sequencing—lighting carriers before transport is fully
upgraded—can turn balance-sheet wins into QoE risks if feedback loops
are slow.
\item \textbf{Culture \& Governance}—
\emph{Culture Eats Strategy for Breakfast} shows that dual-career
ladders have not vanished so much as \emph{rusted}. When those ladders
are repaired—restoring parity between product P\&L and deep technical
stewardship—Tech, Build and Ops clocks re-synchronize and cycle time
compresses. When they are ignored, restructuring
benefits are often drowned out by re-work and hidden copper
shortcuts.
\end{enumerate}

\subsection{6G Design Frontier: Silicon-Paced Complexity}

The LTE → 5G leap was constrained mainly by spectrum physics; the 5G → 6G
leap will be gated by \emph{silicon tempo}. Foundry nodes now shrink every
9–12 months, each bringing new I/O fabrics, chiplets and AI accelerators. A
Tech clock that merely recovers a 4–6 year outlook will still fall behind
unless it also delivers quarterly spec drops and pre-positions at least two
node cycles ahead. A \textbf{dual-resolution planning grid}—five-year
spectrum/infrastructure backbone over-laid with 6–9 month “silicon
sprints”—keeps strategy and execution phase-locked as complexity grows.

\subsection{Open Research \& Engineering Challenges}

\begin{enumerate}[leftmargin=1em]
\item \textbf{Sub‐THz \& JCAS}—Split-1 fronthaul with $T_{\mathrm{e2e}}\le 50\,\mu\mathrm{s}$ and AI-assisted beam discovery.
\item \textbf{AI-native security}—Homomorphic model updates, federated-learning
pipelines and confidential-computing enclaves that defend RIC loops
against poisoning and side channels.
\item \textbf{Energy neutrality}—100× traffic growth under flat power budgets
via photonic/RISC-V accelerators and fine-grain sleep modes.
\item \textbf{Zero-touch assurance}—End-to-end, slice-aware control loops
that bridge radio, transport and core without stalling at data-model
boundaries.
\item \textbf{Multi-RAT Spectrum Sharing (MRSS)}—5G–6G coexistence
through dynamic spectrum sharing that exploits 5G’s forward-compatible
physical layer for efficient refarming.
\item \textbf{Open RAN Everywhere}—Universal 6G RAN–core and gNB–gNB open
interfaces to foster a competitive ecosystem and simplify multivendor
orchestration.
\item \textbf{AI-native Air Interface}—On-device inference for receiver
enhancement, mobility measurement, and cm-level positioning without
central control-plane overheads.
\end{enumerate}

\subsection{Practical Recommendations}

\noindent\textbf{For operators}—Expose transport “truth-tables’’ to RF
planners, link rollout gates to on-air QoE rather than build counts, and pay
distinguished engineers on par with P\&L owners—innovation compounds faster
than interest. \
\textbf{For vendors}—Publish SBOMs, enable remote attestation and keep MAC/PHY
hooks open for xApp experimentation. \
\textbf{For researchers}—Couple accurate RF ray-tracing with full-stack
protocol emulation, explore energy-proportional accelerators, and devise
metrics that co-optimise security, latency and spectral efficiency.

\subsection{Closing Remark}

Virtualized, AI-optimized and security-hardened RANs provide a credible path
to scalable 6G, but they succeed only when culture, timing and infrastructure
move in concert. The key prescription that emerges—\emph{phase-lock the
clocks, sustain deep-tech ladders, surface constraints early, automate only
after the physics are understood}—is neither anti-nor pro-M\&A; it is
simply system science. Organizations that internalise it will translate
capital into capacity and experience; those that neglect it may still build,
but with ever-lower return on spectrum, silicon and talent.

\ifCLASSOPTIONcaptionsoff
  \newpage
\fi

\nocite{*}
\bibliographystyle{IEEEtran}
\bibliography{References}

\begin{thebibliography}{10}
\providecommand{\url}[1]{#1}
\csname url@samestyle\endcsname
\providecommand{\newblock}{\relax}
\providecommand{\bibinfo}[2]{#2}
\providecommand{\BIBentrySTDinterwordspacing}{\spaceskip=0pt\relax}
\providecommand{\BIBentryALTinterwordstretchfactor}{4}
\providecommand{\BIBentryALTinterwordspacing}{\spaceskip=\fontdimen2\font plus
\BIBentryALTinterwordstretchfactor\fontdimen3\font minus \fontdimen4\font\relax}
\providecommand{\BIBforeignlanguage}[2]{{%
\expandafter\ifx\csname l@#1\endcsname\relax
\typeout{** WARNING: IEEEtran.bst: No hyphenation pattern has been}%
\typeout{** loaded for the language `#1'. Using the pattern for}%
\typeout{** the default language instead.}%
\else
\language=\csname l@#1\endcsname
\fi
#2}}
\providecommand{\BIBdecl}{\relax}
\BIBdecl

\bibitem{tmobile_layercake}
\BIBentryALTinterwordspacing
N.~Ray. (2020) T‑mobile’s 5g layer cake explained. Accessed 11 Jun 2025. [Online]. Available: \url{https://www.t-mobile.com/news/network/t-mobile-betty-crocker-5g-layer-cake}
\BIBentrySTDinterwordspacing

\bibitem{verizon_cband}
\BIBentryALTinterwordspacing
{Verizon Communications}. (2024) What is c‑band and what does it mean for 5g? [Online]. Available: \url{https://www.verizon.com/business/resources/articles/s/what-is-c-band-and-what-does-it-mean-for-5g/}
\BIBentrySTDinterwordspacing

\bibitem{Understanding_ORAN}
M.~Polese, L.~Bonati, S.~D’Oro, S.~Basagni, and T.~Melodia, ``Understanding o-ran: Architecture, interfaces, algorithms, security, and research challenges,'' \emph{IEEE Communications Surveys \& Tutorials}, vol.~25, no.~2, pp. 1376--1411, 2023.

\bibitem{oran_spec}
\BIBentryALTinterwordspacing
{O-RAN Alliance}. (2024) O‑ran architecture and acceleration specifications. [Online]. Available: \url{https://www.o-ran.org/technical}
\BIBentrySTDinterwordspacing

\bibitem{never_let_me_down_again}
\BIBentryALTinterwordspacing
B.~Karakoc, N.~F\"{u}rste, D.~Rupprecht, and K.~Kohls, ``Never let me down again: Bidding-down attacks and mitigations in 5g and 4g,'' in \emph{Proceedings of the 16th ACM Conference on Security and Privacy in Wireless and Mobile Networks}, ser. WiSec '23.\hskip 1em plus 0.5em minus 0.4em\relax New York, NY, USA: Association for Computing Machinery, 2023, p. 97–108. [Online]. Available: \url{https://doi.org/10.1145/3558482.3581774}
\BIBentrySTDinterwordspacing

\bibitem{snow_sca}
\BIBentryALTinterwordspacing
H.~Saurabh, A.~Golder, S.~S. Titti, S.~Kundu, C.~Li, A.~Karmakar, and D.~Das, ``Snow-sca: Ml-assisted side-channel attack on snow-v,'' 2024. [Online]. Available: \url{https://arxiv.org/abs/2403.08267}
\BIBentrySTDinterwordspacing

\bibitem{rakuten_bom}
\BIBentryALTinterwordspacing
{Rakuten Mobile}. (2024) Open ran cost structure and deployment economics. Accessed 12 Jun 2025. [Online]. Available: \url{https://corp.mobile.rakuten.co.jp/english/openran-cost}
\BIBentrySTDinterwordspacing

\bibitem{keysight_mimo}
\BIBentryALTinterwordspacing
{Keysight Technologies}. (2020) What is 5g massive mimo? [Online]. Available: \url{https://www.keysight.com/blogs/en/inds/2020/02/19/what-is-5g-massive-mimo}
\BIBentrySTDinterwordspacing

\bibitem{ericsson_ca}
\BIBentryALTinterwordspacing
{Ericsson}, ``Carrier aggregation—boosting 5g performance,'' \emph{Ericsson Technology Review}, 2023. [Online]. Available: \url{https://www.ericsson.com/en/ran/carrier-aggregation}
\BIBentrySTDinterwordspacing

\bibitem{fcc_auction110}
\BIBentryALTinterwordspacing
{Federal Communications Commission}. (2022) Auction 110 (3.45 ghz service) closing public notice. [Online]. Available: \url{https://www.fcc.gov/document/auction-110-closing-public-notice}
\BIBentrySTDinterwordspacing

\bibitem{securing_oran}
J.~Groen, B.~Kim, and K.~Chowdhury, ``The cost of securing o-ran,'' in \emph{ICC 2023 - IEEE International Conference on Communications}, 2023, pp. 5444--5449.

\bibitem{openranpolicy}
\BIBentryALTinterwordspacing
{Open RAN Policy Coalition}. (2023) Policy recommendations for an open and secure ran ecosystem. [Online]. Available: \url{https://www.openranpolicy.org/}
\BIBentrySTDinterwordspacing

\bibitem{cisco_son}
\BIBentryALTinterwordspacing
{Cisco Systems}, ``From self‑optimizing networks to autonomous ran,'' \emph{Cisco Blogs}, 2023. [Online]. Available: \url{https://blogs.cisco.com/tag/self-optimizing-networks}
\BIBentrySTDinterwordspacing

\bibitem{oaic}
P.~S. Upadhyaya, N.~Tripathi, J.~Gaeddert, and J.~H. Reed, ``Open ai cellular (oaic): An open source 5g o-ran testbed for design and testing of ai-based ran management algorithms,'' \emph{IEEE Network}, vol.~37, no.~5, pp. 7--15, 2023.

\bibitem{x5g}
\BIBentryALTinterwordspacing
D.~Villa, I.~Khan, F.~Kaltenberger, N.~Hedberg, R.~S. da~Silva, S.~Maxenti, L.~Bonati, A.~Kelkar, C.~Dick, E.~Baena, J.~M. Jornet, T.~Melodia, M.~Polese, and D.~Koutsonikolas, ``X5g: An open, programmable, multi-vendor, end-to-end, private 5g o-ran testbed with nvidia arc and openairinterface,'' 2025. [Online]. Available: \url{https://arxiv.org/abs/2406.15935}
\BIBentrySTDinterwordspacing

\bibitem{dApps}
\BIBentryALTinterwordspacing
A.~Lacava, L.~Bonati, N.~Mohamadi, R.~Gangula, F.~Kaltenberger, P.~Johari, S.~D'Oro, F.~Cuomo, M.~Polese, and T.~Melodia, ``dapps: Enabling real-time ai-based open ran control,'' 2025. [Online]. Available: \url{https://arxiv.org/abs/2501.16502}
\BIBentrySTDinterwordspacing

\bibitem{lightreading_dss}
\BIBentryALTinterwordspacing
M.~Dano. (2024) The quiet sunset of 5g dynamic spectrum sharing. Accessed 11 Jun 2025. [Online]. Available: \url{https://www.lightreading.com/5g/the-quiet-sunset-of-5g-dynamic-spectrum-sharing}
\BIBentrySTDinterwordspacing

\bibitem{fcc_clearing}
\BIBentryALTinterwordspacing
{Federal Communications Commission}. (2023) C-band satellite relocation benchmarks and deadlines. Accessed 12 Jun 2025. [Online]. Available: \url{https://www.fcc.gov/c-band-clearing}
\BIBentrySTDinterwordspacing

\bibitem{ghoshal2022-wheels-replication}
\BIBentryALTinterwordspacing
M.~Ghoshal, I.~Khan, Z.~J. Kong, P.~Dinh, J.~Meng, Y.~C. Hu, and D.~Koutsonikolas, ``Performance of cellular networks on the wheels,'' in \emph{Proceedings of the 2023 ACM on Internet Measurement Conference}, ser. IMC '23.\hskip 1em plus 0.5em minus 0.4em\relax New York, NY, USA: Association for Computing Machinery, 2023, p. 678–695. [Online]. Available: \url{https://doi.org/10.1145/3618257.3624814}
\BIBentrySTDinterwordspacing

\bibitem{ghoshal2025-wheels-replication}
\BIBentryALTinterwordspacing
M.~Ghoshal, O.~Basit, I.~Khan, Z.~J. Kong, S.~Wang, Y.~Feng, P.~Dinh, Y.~C. Hu, and D.~Koutsonikolas, ``Replication: Performance of cellular networks on the wheels,'' in \emph{Proceedings of the 2025 ACM Internet Measurement Conference (IMC '25)}.\hskip 1em plus 0.5em minus 0.4em\relax Madison, WI, USA: Association for Computing Machinery, 2025, pp. 1--16. [Online]. Available: \url{https://doi.org/10.1145/3730567.3764486}
\BIBentrySTDinterwordspacing

\bibitem{nokia_anyran}
\BIBentryALTinterwordspacing
{Nokia}. (2024) Anyran — cloud ran for every purpose. [Online]. Available: \url{https://www.nokia.com/networks/solutions/cloud-ran}
\BIBentrySTDinterwordspacing

\bibitem{samsung_vran}
\BIBentryALTinterwordspacing
{Samsung Electronics}. (2023) Samsung vran 3.0 delivers next‑level performance. [Online]. Available: \url{https://www.samsung.com/global/business/networks/solutions/vran}
\BIBentrySTDinterwordspacing

\bibitem{snow_v_fpga}
A.~Caforio, F.~Balli, and S.~Banik, ``Melting snow-v: improved lightweight architectures,'' \emph{Journal of Cryptographic Engineering}, vol.~12, pp. 1--21, 04 2022.

\bibitem{nvidia_aerial}
\BIBentryALTinterwordspacing
{NVIDIA}. (2024) Nvidia aerial sdk: Gpu‑accelerated 5g vran. [Online]. Available: \url{https://developer.nvidia.com/aerial}
\BIBentrySTDinterwordspacing

\bibitem{hpc_riscv}
M.~Wei, G.~Yang, and F.~Kong, ``Software implementation and comparison of zuc-256, snow-v, and aes-256 on risc-v platform,'' in \emph{2021 IEEE International Conference on Information Communication and Software Engineering (ICICSE)}, 2021, pp. 56--60.

\bibitem{opensignal_q12025}
\BIBentryALTinterwordspacing
{Opensignal Ltd.} (2025) Mobile network experience report — united states (april 2025). Dataset window: 1 Jan – 31 Mar 2025; accessed 12 Jun 2025. [Online]. Available: \url{https://www.opensignal.com/reports/2025/04/usa/mobile-network-experience}
\BIBentrySTDinterwordspacing

\bibitem{opensignal_q32023}
\BIBentryALTinterwordspacing
------. (2023) 5g experience report—united states (october 2023). Dataset window: 1 July – 30 Sept 2023; accessed 12 Jun 2025. [Online]. Available: \url{https://www.opensignal.com/reports/2023/10/usa/mobile-network-experience}
\BIBentrySTDinterwordspacing

\bibitem{sep_decline}
\BIBentryALTinterwordspacing
{IPlytics GmbH}. (2023) Essential 5g patent families—who leads the race? Operator share fell from 18\,\% (3G) to 3\,\% (5G). [Online]. Available: \url{https://www.iplytics.com/}
\BIBentrySTDinterwordspacing

\bibitem{azhar2025}
A.~Azhar, ``Who’s (not) ready for super intelligence,'' \emph{LinkedIn Pulse}, Jun 2025, accessed: 16 June 2025.

\bibitem{andreessen2024}
M.~Andreessen, ``Incumbents will be nuked; everything will be rebuilt,'' Podcast interview, 2024.

\bibitem{ashby1956}
W.~R. Ashby, \emph{An Introduction to Cybernetics}.\hskip 1em plus 0.5em minus 0.4em\relax Chapman \& Hall, 1956.

\bibitem{forrester1961}
J.~W. Forrester, \emph{Industrial Dynamics}.\hskip 1em plus 0.5em minus 0.4em\relax MIT Press, 1961.

\bibitem{meadows1999}
D.~Meadows, ``Leverage points: Places to intervene in a system,'' Sustainability Institute Report, 1999.

\end{thebibliography}

\end{document}